\documentclass[conference]{IEEEtran}
\IEEEoverridecommandlockouts
\usepackage{cite}
\usepackage{amsmath,amssymb,amsfonts}
\usepackage{algorithmic}
\usepackage{graphicx}
\usepackage{textcomp}
\usepackage{xcolor}
\def\BibTeX{{\rm B\kern-.05em{\sc i\kern-.025em b}\kern-.08em
    T\kern-.1667em\lower.7ex\hbox{E}\kern-.125emX}}
\begin{document}

\title{FinBERT-LSTM: Deep Learning based stock price prediction using News Sentiment Analysis\\

}

\author{\IEEEauthorblockN{1\textsuperscript{st} Shayan Halder}
\IEEEauthorblockA{\textit{Department of Computing Technologies} \\
\textit{SRM Institute of Science and Technology}\\
Kattankulathur, India \\
sc2235@srmist.edu.in}

}

\maketitle

\begin{abstract}
Economy is severely dependent on the stock market. An uptrend usually corresponds to prosperity while a downtrend correlates to recession. Predicting the stock market has thus been a centre of research and experiment for a long time. Being able to predict short term movements in the market enables investors to reap greater returns on their investments. Stock prices are extremely volatile and sensitive to financial market. In this paper we use Deep Learning networks to predict stock prices, assimilating financial, business and technology news articles which present information about the market. First, we create a simple Multilayer Perceptron (MLP) network and then expand into more complex Recurrent Neural Network (RNN) like Long Short Term Memory (LSTM), and finally propose FinBERT-LSTM model, which integrates news article sentiments to predict stock price with greater accuracy by analysing short-term market information. We then train the model on NASDAQ-100 index stock data and New York Times news articles to evaluate the performance of MLP, LSTM, FinBERT-LSTM models using mean absolute error (MAE), mean absolute percentage error (MAPE) and accuracy metrics.
\end{abstract}


\section{Introduction}
Stock market is a very important part of the economy [1].  Public companies acquire capital through the stock market, and this capital funds various R\&D projects to create services, products, employment which helps grow the economy. If a company performs poorly, its stock price plummets and if it performs very well, its price soars. Investors thus need to research the stock market to decide which Investments will lead to profits. 

\indent Predicting stock price is a very complicated task as it does not follow a fixed mathematical equation. It is subjected to change at any moment of time depending on multiple factors like inflation, geopolitical conflicts, etc [2]. However, we can make fairly accurate short-term predictions by learning the patterns in data. Machine Learning techniques have been used to solve the short-term prediction problem and people have had success in predicting fairly accurate figures [3]. Artificial Neural Networks (ANN) can learn various trends from previous stock price data and use that knowledge to make future predictions. News articles are valuable data in stock market analysis [4]. News articles contain information about the market, negative articles are linked with poor performance of company and positive articles with good performance. Hence news articles can be studied to understand the trend of a stock.

\indent Sentiment analysis can be used to gain information from news articles. Sentiment analysis is understanding the reaction of people to news articles using Natural Language Processing (NLP) techniques. Sentiments can capture general public’s feedback to news articles.  A news article talking about profit, acquisition has positive appeal to people and thus drive its stock price up, while an article talking about layoffs, bankruptcy has a negative appeal and causes its stock price to go down. Hence, understanding sentiment of news articles help us to gain insight into a company’s performance and thus enable us to make predictions on its future stock price.

\indent We use FinBERT, a pre-trained NLP model to analyze sentiment of financial text [5]. FinBERT is built by further training and fine-tuning BERT, a state-of-the-art NLP model [9]. FinBERT returns positivity, negativity, neutrality scores which tell us how positive or negative news articles are on a particular day.

\indent Deep Learning models have been very successful in time series prediction given abundant data to learn from [6]. We thus use Deep Learning for this stock prediction challenge. We begin with a very simple Multilayer Perceptron (MLP) model. We evaluate the performance of this model on our NASDAQ-100 index data. Then we explore more complex model like Recurrent Neural Network (RNN) which is powerful for sequence where the order of the data matters such as stock price over a period of time. We build a Long Short Term Memory (LSTM) model which is an RNN architecture that solves the problem of vanishing gradients. We compare its performance with the MLP we built previously.

\indent Finally, we construct a FinBERT-LSTM model which combines news sentiment from FinBERT with daily stock price data to make future prediction, and compare its performance with the MLP and vanilla LSTM models. 

\indent In section 2 of this paper, we explore the stock price models, news sentiment analysis model. In section 3 we describe our data, model architecture. We present our results in section 4 and conclude the paper in section 5.

\section{Models}

In this section we first introduce a Multilayer Perceptron (MLP) model, then Long Short Term Memory (LSTM) model and finally the FinBERT model.

\subsection{MLP network}

Multilayer Perceptron is a simple fully connected Feed Forward Artificial Neural Network (ANN). It has one or more hidden layers between an input layer and an output layer. The weights and biases of the connections are adjusted through backpropagation [7].

\begin{figure}[!htb]
\centerline{\includegraphics[width = 260pt]{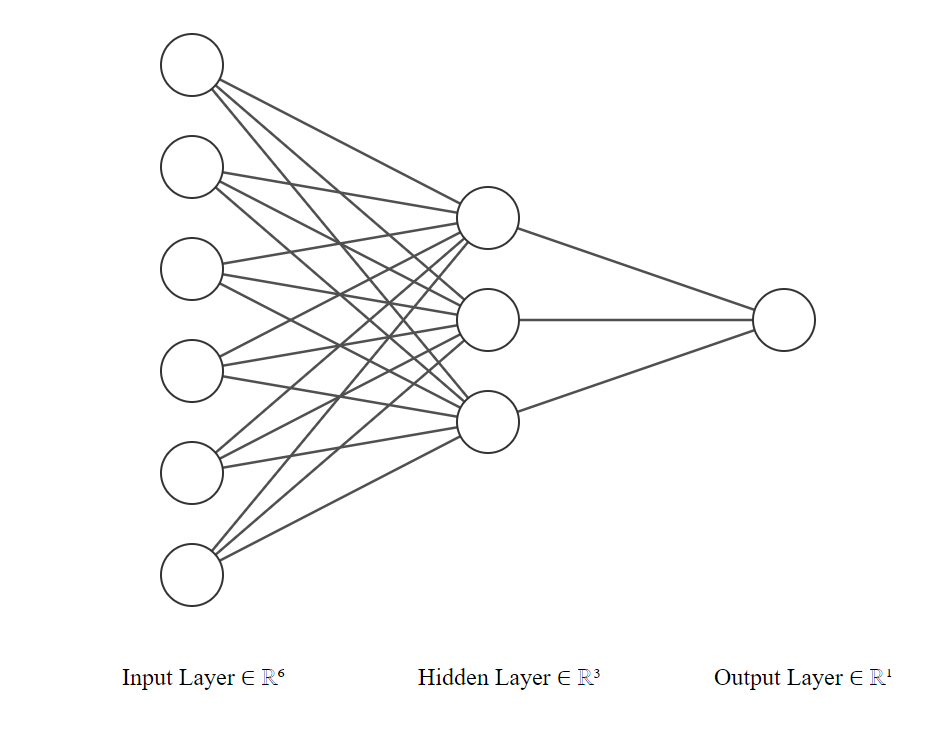}}
\caption{Structure of a 3 layer MLP}
\label{fig}
\end{figure}

Above is a simple 3 layer MLP model. The input layer has 6 neurons, the hidden layer has 3 neurons and the output layer has 1 neuron. Each neuron is represented by a circle and the weights between them are represented by straight lines. These weights and biases are adjusted during backpropagation where an optimizer such as Adam or Stochastic Gradient Descent is used to minimize a certain cost function such as mean squared error (MSE).

\subsection{LSTM network}

Recurrent Neural Networks (RNN) are widely used for applications where the sequence matters [8]. The memory cells of RNN allows the network to remember previous information. Long Short Term Memory (LSTM) is a Recurrent Neural Network (RNN) that addresses the problem of vanishing gradients of conventional RNN. LSTM allows a model to remember previous information for longer period of time. An LSTM cell consists of 3 gates: Input gate,  Output gate, Forget gate.

\begin{figure}[!htb]
\centerline{\includegraphics[width = 260pt]{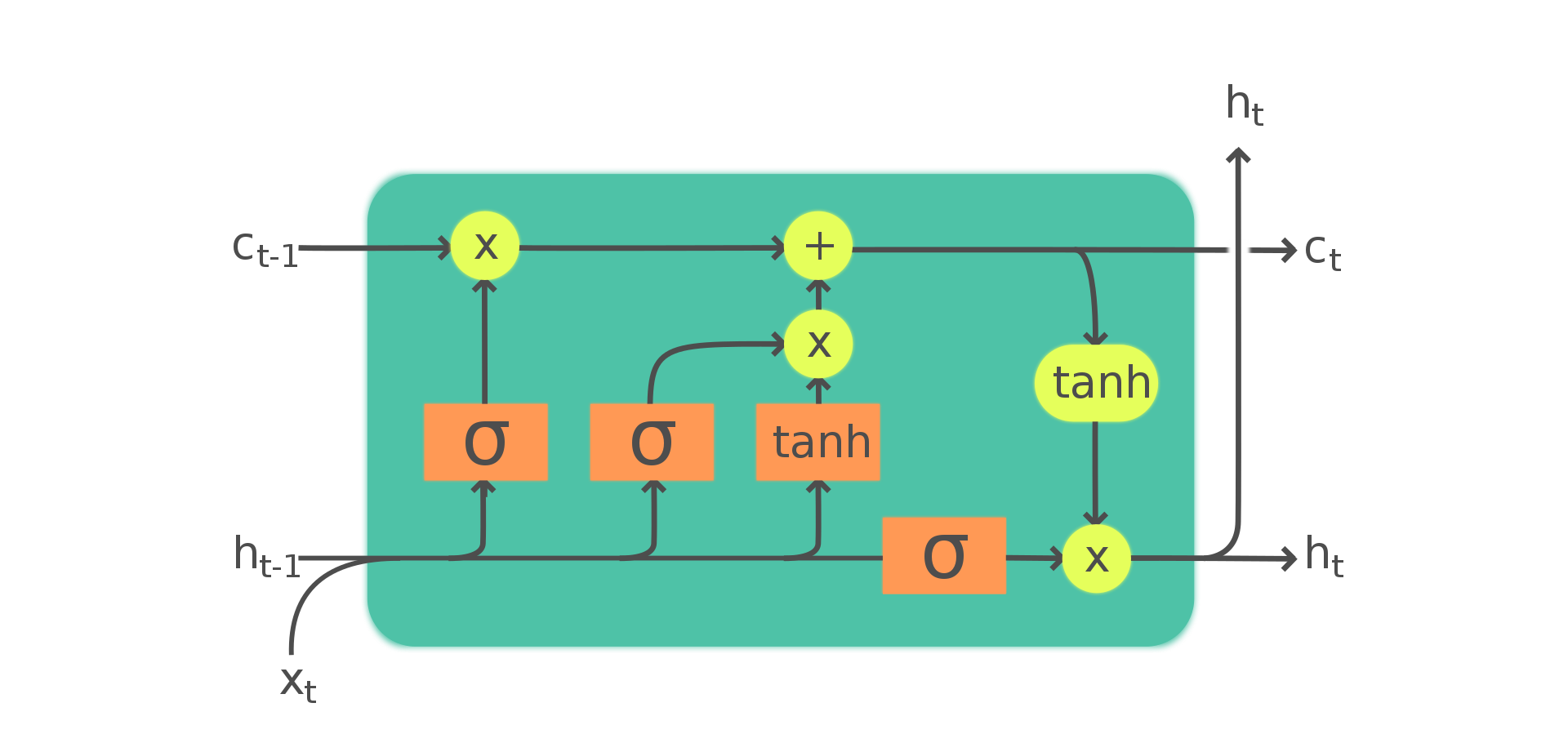}}
\caption{LSTM Cell [10]}
\label{fig}
\end{figure}

\begin{itemize}
  \vspace{3mm}
  \item \textbf{Input gate} – The input gate controls the flow of input activations into the memory cell. It is given by
  
  \[i_t= \sigma(W_i \times [h_{t-1},X_t ]+b_i)\]
  
  where \(i_t\) is the output of the input gate at time step \(t\), \(W_i\) is the weight of input gate, \(h_{t-1}\) is the hidden state at time step \(t-1\), \(X_t\) is the input at time step \(t\) and \(b_i\) is the bias of input gate.
  
  \[\hat{c} = tanh(W_c \times [h_{t-1}, X_t] + b_c)\]
  
  where \(\hat c\) is the candidate value to be added to output at time step \(t\), \(W_c\) is the weight of candidate value, \(b_c\) is the bias of candidate value.
  \vspace{3mm}
  \item \textbf{Forget gate} - The forget gate controls which information to remember and which information to forget in memory. It is given by
  
  \[f_t = \sigma(W_f \times [h_{t-1}, X_t] + b_f)\]
  
  where \(f_t\) is the output of the forget gate at time step \(t\), \(W_f\) is the weight of forget gate, \(b_f\) is the bias of forget gate.
  \vspace{3mm}
  \item \textbf{Output gate} - The output gate controls the information being output from memory cell. It is given by
  
  \[o_t = \sigma(W_o \times [h_{t-1}, X_t] + b_o)\]
  
  where \(o_t\) is the output of the output gate at time step t, \(W_o\) is the weight of output gate, \(b_o\) is the bias of the output gate.
  
  \[c_t = f_t \times c_{t-1} + i_t \times \hat c_t\]
  
  where \(c_t\) is the new cell state at time step \(t\).
  
  \[h_t = o_t \times tanh(c_t)\]
  
  where \(h_t\) is the new hidden state at time step \(t\).
  
\end{itemize}

\subsection{FinBERT model}

Textual information from news can be very helpful to predict the short-term movement of stock. Sentiment analysis of news titles can be a useful feature to determine the immediate performance of a stock. Bidirectional Encoder Representations from Transformers (BERT) is a state-of-the-art Machine Learning model for Natural Language Processing (NLP) developed by Google [9]. BERT is bidirectionally trained, which helps in understanding context better from both left and right side simultaneously. BERT uses transformers which learns the contextual relationship between words in a text. BERT is pre-trained on two tasks: Masked Language Modelling (MLM), Next Sentence Prediction (NSP). The original BERT paper provides a more detailed description of the model architecture, pre-training and fine-tuning.

\begin{figure}[!htb]
\centerline{\includegraphics[width = 280pt]{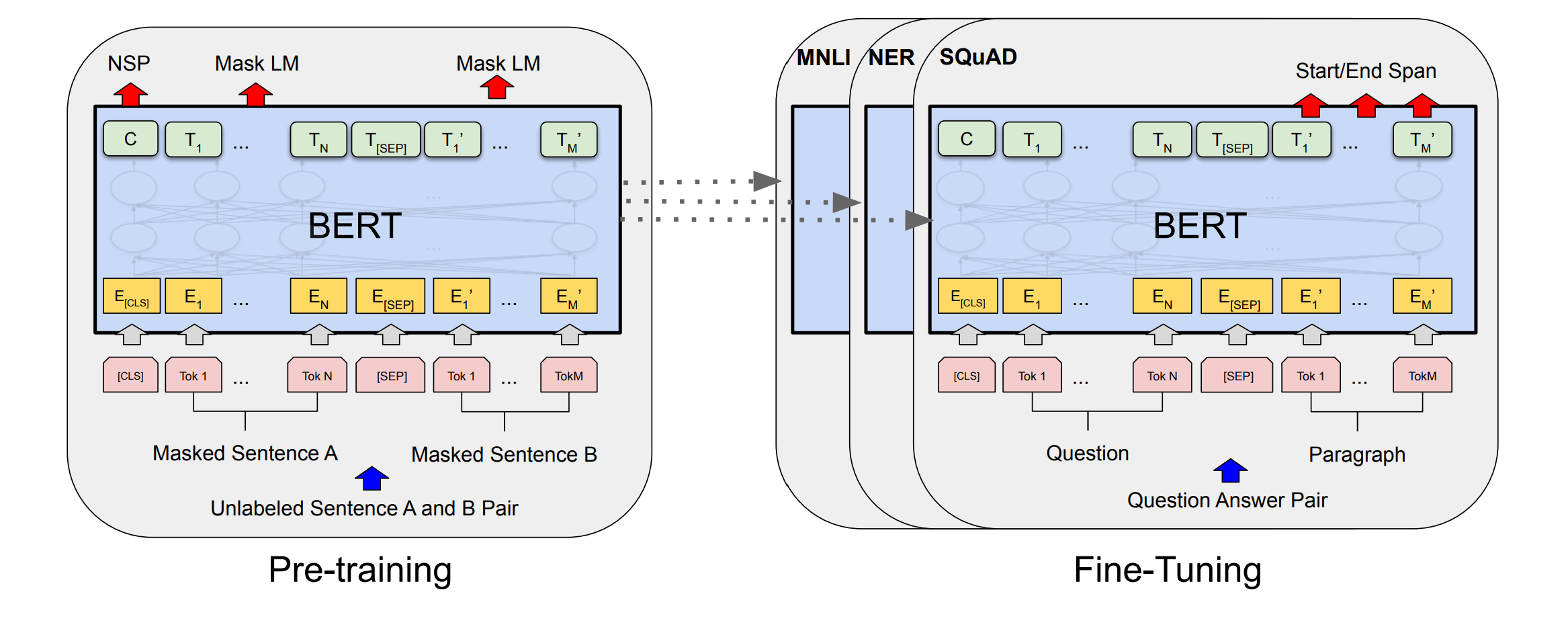}}
\caption{Pre-training and fine-tuning procedures for BERT. [CLS] is a special symbol added in front of every input example, and [SEP] is a special separator token [9]}
\label{fig}
\end{figure}

We use FinBERT, a pre-trained BERT model fine-tuned for financial sentiment classification [5]. It is open source under Apache License, Version 2.0. FinBERT analyses strings and outputs a score of 0 to 1 and the sentiment label: positive, negative, neutral. Higher score demonstrates higher confidence in the corresponding label. 

\section{Model training}

\subsection{Data preprocessing}

We collect 2 types of data for this project. We use Yahoo Finance API to download NASDAQ-100 index stock price data in the time period \(01/10/2020\) to \(30/09/2022\). In this project we use daily closing price data for training and prediction. 

\indent

For news sentiment analysis we use news articles from New York Times. NY Times is an American daily newspaper based in New York City. It is one of the most circulated newspapers in US and covers a wide range of topics including Finance, Business, Technology which is our requirement in this project. We use the NY Times API to collect up to 10 news articles in the domain of finance, business and technology of every single day in the period \(01/10/2020\) to \(30/09/2022\). We then pass a list of news headings on a particular day to the FinBERT model to compute the sentiment on that day.

\indent

We create a rolling window of 10 days, we feed our neural network data of \(10\) days and let it predict stock price of the \(11^{th}\) day. If we have data of \(n\) days, then to predict the stock price of \(i^{th}\) day we feed it data of \(i - 10\) to \(i - 1\) days. We repeat the process \(\forall i \in (11, n)\) by shifting the window including 1 new data point at a time.

\indent 

We use 85\% of our data for training and use the remaining 15\% for testing to evaluate the performance of our model.

\subsection{Normalization}

We normalize our stock price data. Reforming the data on a common scale helps our model learn patterns and thus perform better. We use min-max normalization on our data. It is given by

\[X^n_t = \frac{X_t - min(X_t)}{max(X_t) - min(X_t)}\]

where \(X^n_t\)t is the data point \(X_t\) after normalization. inverse-normalization can be done by

\[X_t = X^n_t \times [max(X_t) - min(X_t)] + min(X_t)\]

\subsection{Model architecture}

In the MLP model we have 5 layers: 1 input layer, 3 hidden layers and 1 output layer. The input layer accepts input of the shape \((10, 1)\). The hidden layers are dense layers and have ReLU activation function. Let number of hidden layers be \(l\), here \(l = 3\). For \(l = 1\), we have 50 neurons, for \(l = 2\) we have 30 neurons and for \(l = 3\) we have 20 neurons, for output layer we have 1 neuron. We also have 3 dropout layers in between (with dropout rates \(0.1, 0.05, 0.01\) respectively) to prevent overfitting.

\indent 

We use Mean Squared Error (MSE) loss function, which is the sum of the squared distances between target variable and predicted value and train the model using Adam optimizer with learning rate \(0.01\).

\indent 

We then construct an LSTM network of 5 layers: 1 input, 3 hidden, 1 output. The input layer accepts input of the shape \((10, 1)\).  For the hidden layers we use LSTM layers with “tanh” activation function. For layer \(l = 1\), we have 50 neurons, 30 neurons in \(l = 2\), 20 neurons in \(l = 3\) and 1 neuron in the output layer. We add 3 dropout layers in between them (with dropout rate \(0.1, 0.05, 0.01\) respectively) to avoid overfitting.

\indent 

We train the LSTM network using MSE loss function and Adam optimizer having \(0.02\) learning rate.

\indent

Finally, we combine stock data and news sentiment together to from the FinBERT model. We also use 5 layers: 1 input, 3 hidden, 1 output layers in this model. The input layer in FinBERT model accept input of shape \((11, 1)\), as we combine closing stock price of 10 days and news sentiment on that day resulting in 11 features. The LSTM hidden layers use “tanh” activation function and layer \(l = 1\) has 70 neurons, \(l = 2\) has 30 neurons, \(l = 3\) has 10 neurons and the output layer has 1 neuron. We do not add any dropout layer as we did not notice any overfitting during training.

\indent 

We use MSE loss function and Adam optimizer with a learning rate of \(0.02\).

\section{Performance}

In this section we evaluate the performance of all 3 models we built and compare them. We use three metrics to assess the performance:

\begin{itemize}
  \vspace{3mm}
  \item \textbf{Mean Absolute Error} – Mean Absolute Error (MAE) is the mean of the distance between the target variable and predicted value. MAE can be computed by
  
  \[\textrm{MAE} = \frac{\Sigma^n_{i = 1}|\hat y_i - y_i|}{n}\]
  
  where \(\hat y_i\) is the predicted value for training example \(i\), \(y_i\) is the target variable for training example \(i\).
  \vspace{3mm}
  \item \textbf{Mean Absolute Percentage Error} – Mean Absolute Percentage Error (MAPE) is the mean of the distance between the target variable and the predicted value as a fraction of the target variable. MAPE can be computer by
  
  \[\textrm{MAPE} = \frac{1}{n} \Sigma^n_{i = 1} |\frac{\hat y_i - y_i}{y_i}|\]
  
  \item \textbf{Accuracy} – Accuracy can be calculated by
  \vspace{3mm}
  \[\textrm{Accuracy} = 1 - \textrm{MAPE}\]
  
\end{itemize}

\begin{figure}[!htb]
\centerline{\includegraphics[width = 310pt]{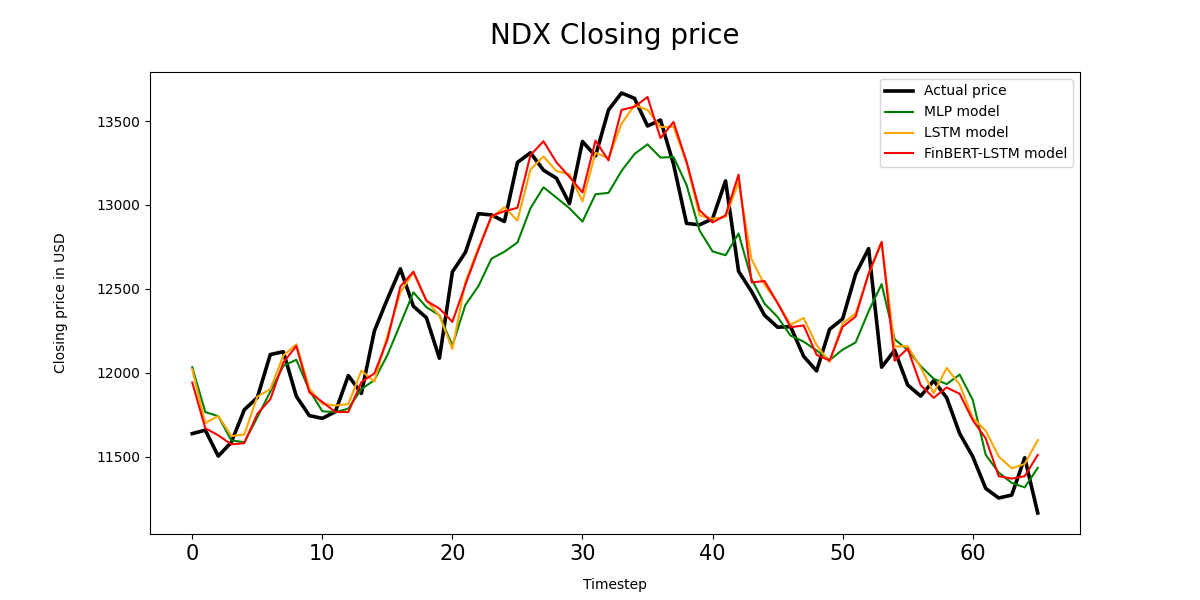}}
\caption{Prediction results}
\label{fig}
\end{figure}

From the plot in Figure 4 we can see the predictions of FinBERT-LSTM and LSTM are closer to the actual price than the MLP model. 

\begin{table}[!ht]
    \centering
    \caption{Performance on NASDAQ-100 index}
    \begin{tabular}{|c|c|c|c|}
    \hline
        \textbf{Models} & \textbf{MAE} & \textbf{MAPE} & \textbf{Accuracy} \\ \hline
        \textbf{MLP} & 218.32973474 & 0.01767204122 & 0.98232795877 \\ \hline
        \textbf{LSTM} & 180.58083886 & 0.01456811176 & 0.98543188823 \\ \hline
        \textbf{FinBERT-LSTM} & 174.94284259 & 0.01409574846 & 0.98590425153 \\ \hline
    \end{tabular}
\end{table}

From the metric we can notice that our FinBERT-LSTM model has a 20.2370\% MAPE and 0.3640\% accuracy performance improvement over the MLP model. FinBERT-LSTM also performs better than the vanilla LSTM model, achieving 3.2424\% MAPE and 0.0479\% accuracy performance improvement. Adding news sentiment helped the model to identify stock patterns better.

\indent 

The results are obtained after fine-tuning each model and running multiple trials on the best performing architecture. We ran 1000 trials on each model to mitigate randomness.

\section{Conclusion}

Predicting stock price is a very challenging task. In this paper we explored different Deep Learning techniques to predict stock price. First, we created a Multilayer Perceptron (MLP) network. Then we discussed about Long Short Term Memory (LSTM) network, which is a Recurrent Neural Network (RNN). Finally, we proposed FinBERT-LSTM model. This model incorporates news sentiments into an LSTM network to make predictions. We used FinBERT, a pretrained BERT model fine-tuned for financial sentiment to compute the sentiment on each day. Experiments and testing on NASDAQ-100 index showed the FinBERT-LSTM model outperform all other models. Prediction results show that adding news sentiment to the training data can help the model learn better about the stock market patterns and thus make more robust predictions.

\end{document}